# Theory of oxidation/reduction-induced chromium ion valence transformations in Cr,Ca:YAG crystals

M. Sinder, Z. Burshtein and J. Pelleg

Department of Materials Engineering, Ben-Gurion University of the Negev, P.O. Box 653,

Beer-Sheva 84105, Israel

## Abstract

In their paper [Opt. Mater. **24**, 333, 2003], Feldman et al conducted an experimental study of the dynamics of chromium ion valence transformations in Cr,Ca:YAG crystals among the trivalent $Cr^{3+}$ state, and two tetravalent $Cr^{4+}$ ones, of octahedral and tetrahedral coordination. The temperatures used ranged between ~800 and $1,000^{\circ}C$. The basic effects are the transition of $Cr^{3+}$ into $Cr^{4+}$ under high-temperature annealing in an oxidizing atmosphere, and the reverse transition under a reducing atmosphere, or in vacuum. In the present theory, we interpret the processes by oxygen-vacancy diffusion in the bulk of the YAG. The quasi-chemical reaction $V_O^{2-} + Cr^{4+} \Leftrightarrow Cr^{3+} / V_O^{-}$ between the chromium ions and the vacancies $V_O^{2-}$ is responsible for the valence transformations. The external oxygen pressure determines the equilibrium oxygen vacancy concentration at the surface, which, after a sufficiently long time, extends by diffusion throughout the entire sample. The $Cr^{4+}$ equilibrium concentration is proportional to the square-root of the oxygen partial pressure at low pressures, and saturates at high pressures, when the entire population of $Cr^{3+}$ ions, adjacent to $Ca^{2+}$ ions, has oxidized. Dynamical analysis provides profiles of the $Cr^{3+}$ and $Cr^{4+}$ concentrations in the crystal during oxidation and during reduction. Reaction rate profiles are also calculated, establishing the reaction front position and width. Reduction is always chracterized by two well-defined zones, each near the sample opposite surface, where the $Cr^{4+} \rightarrow Cr^{3+}$ transition reaction takes place. The zone boundaries move continually towards the sample center. During oxidation, oxygen vacancies diffuse towards the surface. No reaction front forms. For sufficiently long times the $Cr^{3+} \rightarrow Cr^{4+}$ transition reaction progresses throughout the entire sample. A comparison with existing experimental results on the integrated $Cr^{4+}$ concentration as a function of time during oxidation [Opt. Mater. **24**, 333, 2003] shows reasonable agreement.

# I. Introduction

In recent years, a special interest arose concerning chromium in its tetravalent state. Different $Cr^{4+}$-doped oxide crystals were studied as both laser gain materials for the near IR (1.3–1.5 μm) [1-4], and as saturable absorbers used for passive Q-switching of IR lasers in the 1.0–1.1 μm range [3,5-7]. In their paper, Feldman et al [8] studied the dynamics of chromium ion valence transformations in Cr,Ca:YAG crystals ($Cr$**,**$Ca$ **:** $Y_3Al_5O_{12}$) among the trivalent $Cr^{3+}$ state, and two tetravalent $Cr^{4+}$ ones, of octahedral and tetrahedral coordination. The tetravalent $Cr^{4+}$ concentrations were typically of the order of $10^{18}$ $cm^{-3}$, while the total chromium concentration was at least an order of magnitude larger. The transformations were induced by high-temperature oxidation or reduction of the crystals in oxygen ambient, or under vacuum, respectively. The role of divalent calcium ions $Ca^{2+}$ in the composition, is charge compensation, which allows the formation of tetravalent chromium ions. Yet in the crystal, it appears that only a small fraction of the chromium atoms, 1-5%, actually existed in the tetravalent state after prolonged oxidation, even when the Ca concentration exceeded that of the $Cr^{4+}$. Sugimoto et al [9] suggested, that a considerable proportion of the $Ca^{2+}$ ions, are charge-compensated by oxygen vacancies.

Okhrimchuk et al [10-12] considered the involvement of singly-negative oxygen vacancies $V_O^-$, and free holes, in the oxydation/reduction reactions. The free holes, diffusing through the crystal, are responsible for the valence transformation of octahedral chromium $Cr^{3+} \Leftrightarrow Cr^{4+}$. A thermally activated position exchange between octahedral $Cr^{4+}$ and tetrahedral $Al^{3+}$ allows the appearance of tetrahedral $Cr^{4+}$. Various studies of high-temperature conductivity in sintered or crystalline YAG, doped with different ions, established the role of free electrons and holes in the electrical conduction; notably, the works of Rotman et al [13 and inherent references] on crystalline YAG, and of Schuh et al [14] on sintered YAG. The electronic conductivity could be observed at high temperatures, yet it reduced exponentially on reduction of temperature (for example, Figs 6, 7 in [13]). Both ionic and electronic conductivities practically diminished below approximately 1100$^\circ$C.

The theory developed in the present work is based on the choice of $V_O^{2-}$ as the specific vacancy type involved in the oxidation reduction process (i.e., no free electrons nor holes are assumed to be present). An important advantage of the theory is that it points out experimental procedures that could support or contradict this choice.

## II. Model

For the benefit of the readers we describe our model in detail. We consider that during reduction of the Cr,Ca:YAG crystals, free oxygen $\frac{1}{2}O_2$ evaporates from the crystal surface. Naturally, oxygen vacancies are created at the crystal surface, which then diffuse into the bulk. During oxidation, the process is reversed. Since lattice oxygen exists as a doubly-negative ion, the evaporation of a neutral atom leaves behind a free lattice site, plus two electrons. A crucial point in construction of our theoretical model is the charge state of the vacancy, namely, the fate of the two said electrons. In priciple, three options may be considered (for example, see p. 41, Table 22.2 in ref [15]): in the first one, the two electrons remain bound to the "free lattice-site", thus forming a $V_O^{2-}$-type vacancy, which is in fact an electrically neutral species. In the second option, one electron turns free, the other remains bound to the free lattice site, and forms a $V_O^-$-type vacancy. In the third option, both electrons turn free, and the free lattice site remains bare - $V_O$.

Feldman et al [8] conducted most of their experiments between 800 and 1000$^\circ$C. Based on conclusions from the works of Rotman et al [13] and Schuh et al [14], neither free ions nor free charge carriers (electrons or holes) may play a significant role in electrical conduction, hence in diffusion, in the said temperature range. Thus in our model, we consider only the involvment of $V_O^{2-}$-type vacancies in the diffusion process. The oxygen evaporation at the YAG surface is then described by the chemical reaction

$$2O^{2-} \Leftrightarrow 2V_O^{2-} + O_2(gas). \tag{1}$$

The original role of $Ca^{2+}$ ions in the $Ca,Cr:YAG$ crystals was charge compensation, to allow the formation of $Cr^{4+}$ ions; the latter are the ones that take part in the reducton/oxidation process [8]. In our model we ignore the involvment in the oxydation/reduction process of such $Ca^{2+}$ ions that may be charge-conpensated by oxygen vacancies [9]. This is supported by the fact that $Cr^{4+}$ concentration exhibits saturation after prolonged oxidation [8]. In fact, the $Ca^{2+}$ and $Cr^{4+}$ ions may reside on the same unit cell, or alternatively, their distance may exhibit a broad distribution (see e.g. ref [15], pp. 242-244). Based on the different ionic radii, and some experimental results, Okhrimchuk et al consider that the existance of $Cr^{4+}-Ca^{2+}$ complex is very probable [12].

In our model we do not consider any significant diffusivity of the $Ca^{2+}$, nor of the $Cr^{4+}$ ions, namely the distance distribution is irrelevant for any of our following discussions. Feldman et al [8] as well as others [9,10, 12] also showed, that a large concentration of $Cr^{3+}$ ions, not compensated by the $Ca^{2+}$ to form $Cr^{4+}$, also existed in the $Ca,Cr:YAG$ crystal. They, however, do not participate in the oxidation/reduction process, and will be overlooked in our theory.

Thus in our model, double-negative oxygen vacancies $V_O^{2-}$ diffuse through the lattice. The net result of $V_O^{2-}$ vacancy diffusion is the exchange of a neutral oxygen atom between its original position and the free site (vacancy). Therefore, no net electrical charge movement occurrs. When the $V_O^{2-}$ vacancy approaches the $Cr^{4+}$, a reaction occurs: one electron is transferred from the $V_O^{2-}$ to the $Cr^{4+}$ to form a stable, immobile $Cr^{3+}/V_O^-$ couple. The stability is due to a Coulomb attraction between the $Cr^{3+}$ end of this couple, which is negative compared to its former state as $Cr^{4+}$, and the $V_O^-$ end, which is positive compared to its former state as $V_O^{2-}$. This reaction is described by

$$V_O^{2-} + Cr^{4+} \Leftrightarrow Cr^{3+}/V_O^- \quad . \tag{2}$$

For brevity, we simplify the notation by writing $Cr^{3+}$ instead of $Cr^{3+}/V_O^-$. One should, however, be careful not to confuse Ca-compensated $Cr^{3+}$ with the uncompensated $Cr^{3+}$.

The $Cr^{4+}$ species includes both its octahedrally and tetrahedrally coordinated states. This exchange of coordination is of prime importance for technological applications of $Ca,Cr:YAG$ - made components. This material is used for either passive Q-switches in laser resonance cavities, or as a laser gain media; it is the $Cr^{4+}$ in its tetrahedral coordination state, which is active in both applications [4,5]. Exchange from an octahedral to a tetrahedral state most likely involves the exchange of positions with a lattice $Al^{3+}$ ion; it requires a considereable activation energy (probably several electron volts, fig. 10 in Ref. [8]). One may then safely assume that the process takes place within the same unit cell, and no long-range diffusion needs to be is assumed. As far as oxidation/reduction reactions are considered, the coordination state of $Cr^{4+}$ is thus irrelevant.

## III. Thermodynamic analysis

Applying the law of mass action to Eq. (1) yields

$$\frac{[V_O^{2-}]^2[O_2]}{[O^{2-}]^2}=K_1 \quad ,$$

(3)

Where $K_1$ is the reaction constant of Eq. (1). Under ordinary experimental conditions, the density of oxygen vacancies always remains extremely small compared to that of the lattice oxygen; thus $[O^{2-}]$ may be assumed constant throughout. In the ideal-gas limit, the density of oxygen atoms in the ambient atmosphere is given by $[O_2]=(P/k_BT)$, where $P$ is the partial oxygen pressure, $k_B$ is Boltzmann's constant, and $T$ is the absolute temperature. Then, one may write

$$[V_O^{2-}]=[O^{2-}]\sqrt{\frac{K_1k_BT}{P}}.$$

(4)

Clearly, the condition $[V_O^{2-}]\ll[O^{2-}]$ requires the use of a sufficiently high oxygen pressure: $P \gg K_1k_BT$. Equation (4) would not be valid if a lower pressure is used! Still, to keep expressions simple, we assume hereafter that the condition on the pressure is satisfied, namely, that equation (4) is always valid.

Applying the law of mass action to Eq. (2) yields

$$\frac{[V_O^{2-}][Cr^{4+}]}{[Cr^{3+}]}=K_2 \quad ,$$

(5)

Where $K_2$ is the reaction constant of Eq. (2). Inserting (4) into (5) one obtains

$$\frac{[Cr^{4+}]}{[Cr^{3+}]}=[O^{2-}]\frac{K_2}{\sqrt{K_1k_BT}}\sqrt{P}\equiv S\sqrt{P} \quad ,$$

(6)

where $S\equiv[O^{2-}]K_2/\sqrt{K_1k_BT}$ is a constant at a set temperature.

Equation (6) indicates that the ratio of tetravalent $Cr^{4+}$ to trivalent $Cr^{3+}$ chromium concentrations is proportional to the square root of the ambient oxygen partial pressure. The sum of tetravalent and trivalent chromium ions is a constant. Specifically,

$$[Cr^{4+}] + [Cr^{3+}] = [Cr^{4+}]_{max} \; . \tag{7}$$

Inserting (7) into (6), and solving for $[Cr^{4+}]$ yields

$$[Cr^{4+}] = [Cr^{4+}]_{max} \frac{S\sqrt{P}}{1 + S\sqrt{P}} \; . \tag{8}$$

A plot of $[Cr^{4+}]$ as a function of $\sqrt{P}$ starts linearly with a slope of $[Cr^{4+}]_{max} \cdot S$, reaches half maximum at a pressure satisfying $\sqrt{P} = 1/S$, and approaches asymptotically a saturation value $[Cr^{4+}]_{max}$ for high $\sqrt{P}$. An experimental study employing the measurement of teravalent chromium ion concentration $[Cr^{4+}]$, as a function of $\sqrt{P}$, is straightforward through optical absorption, since the absorption spectrum of $[Cr^{4+}]$ is already well known [8]. Particularly, experimental values of $K_2/\sqrt{K_1}$ would be obtained, as well as their temperature dependence, by carrying out the same measurements and analyses at different temperatures. One further understands that under low pressure ($P \rightarrow 0$), the thermal-equilibrium concentration, $[Cr^{4+}]$, diminishes.

## IV. Dynamic analysis

For brevity we denote the free vacancy concentration $[V_O^{2-}] \equiv V$, and $[Cr^{4+}]_{max} \equiv C$. Using these notations, Eq. (8) can be written as (see also Eq. (5))

$$[Cr^{4+}] = C \frac{1}{1 + V/K_2} \; . \tag{9}$$

Similarly, one obtains

$$[Cr^{3+}] = C \frac{V/K_2}{1 + V/K_2} \; . \tag{10}$$

The oxygen doubly-negative vacancy concentration $V$ itself is a dynamic variable, determined primarily by its in-, or out-diffusion through the crystal surface. This diffusion is accompanied by a reaction according to Eq. (2). For simplicity we assume a one-dimensional geometry for diffusion, by forming the crystal as a plane-parallel plate. Thus $V = V(x, t)$, where $x$ is the position inside the sample, and $t$ is the time. Both diffusion and reaction, however, govern the concentration of the different species involved:

$$\frac{\partial V(x,t)}{\partial t} = D \frac{\partial^2 V(x,t)}{\partial x^2} - R(x,t) \ ; \tag{11}$$

$$\frac{\partial [Cr^{4+}]}{\partial t} = -R(x,t) \ \ ; \tag{12}$$

$$\frac{\partial [Cr^{3+}]}{\partial t} = +R(x,t) \ \ , \tag{13}$$

where the reaction rate $R(x,t)$ is the net between the two inverse processes of Eq. (2):

$$R(x,t) \equiv K_R \left\{ V \, [Cr^{4+}] - K_2 [Cr^{3+}] \right\} \ , \tag{14}$$

with $K_R$ being the kinetic reaction constant for the process described in Eq. (2). Note that $[Cr^{4+}]$ and $[Cr^{3+}]$ in Eqs.(12) and (13) are also functions of $x$ and $t$.

Summing (11) and (13) yields

$$\frac{\partial}{\partial t} \left\{ V + [Cr^{3+}] \right\} = D \frac{\partial^2 V}{\partial x^2} \ . \tag{15}$$

As a zero-order approximation for solving the diffusion equation we assume that local quasi steady-state conditions are established, namely that $R(x,t) \cong 0$ throughout the entire process [16]. One may then use the result given in (10) as an estimate for the local concentration of $[Cr^{3+}]$. Then,

$$\left\{ 1 + \frac{C/K_2}{\left(1 + V/K_2\right)^2} \right\} \frac{\partial V}{\partial t} = D \frac{\partial^2 V}{\partial x^2} \quad . \tag{16}$$

Equation (16) is a diffusion equation for the vacancy concentration $V$. The additive unity

term, "1", in the brackets on the left handside of Eq. (16) relates to the variation of the free

vacancy concentration. The second term relates to the variation in the concentration of the

bound oxygen vacancy, namely $Cr^{3+}/V_O^-$. Once Eq. (16) is solved, $V(x,t)$ is obtained, and

the $[Cr^{3+}]$ and $[Cr^{4+}]$ concentration profiles and temporal variations immediately follow

(Eqs. (9), (10)). The use of Eqs. (9) and (10) in fact completes a zero-order iteration for estimating $[Cr^{3+}]$ and $[Cr^{4+}]$, which is based on the assumption that $R(x,t) = 0$. A first order approximation for $R(x,t)$ itself may be obtained by using either ~~of~~ equations (10),

(11), or (12) [17,18,19]. The integral reaction rate across the entire sample $\mathbf{R}(t) \equiv \int R(x,t)dx$

follows directly, which may be compared with existing experimental results [8].

   Eq. (16) describes, in fact, the diffusion of oxygen vacancies throughout the bulk. When the

vacancy concentration grossly exceeds the reaction constant $K_2$, namely $V/K_2 \gg 1$, most

chromium atoms reside as trivalent $Cr^{3+}$. When this condition is reversed , namely

$V/K_2 \ll 1$, most chromium atoms reside in their tetravalent $Cr^{4+}$ state (see Eqs. (9),(10)).

Change in vacancy concentration, for example from $V/K_2 \gg 1$ to $V/K_2 \ll 1$ induces

transformation of most chromium atoms from $Cr^{3+}$ to $Cr^{4+}$; change from $V/K_2 \ll 1$ to

$V/K_2 \gg 1$ induces the inverse transformation, of most chromium atoms from $Cr^{4+}$ to

$Cr^{3+}$ [8]. Thus $V \sim K_2$ plays the role of a critical vacancy concentration, controlling the

transformation between the two chromium states. In solving our equations we refer to initial conditions, where all chromium atoms reside as either $Cr^{3+}$ or alternatively as $Cr^{4+}$. A temporally constant vacancy surface concentration $V_s$ will always be assumed, starting at $t = 0$, which is defined by the ambient oxygen partial pressure (Eq. 4). Constancy of $V_s$ implies that the vacancy surface-crossing velocity is infinitely large [20,21]. We are interested in studying valence transformations of chromium ions under different initial conditions. Specifically $V_s \gg K_2$ and $V_0 \ll K_2$, or alternatively, $V_s \ll K_2$ and $V_0 \gg K_2$, where $V_0$ is the (spatially constant) bulk vacancy concentration at $t = 0$.

   We turn now to solving Eq. (16) in the frame of the latter initial and boundary conditions, for various ratios of $C/K_2$ for cases of vacancy in-diffusion $V_s \gg V_0$, or out-diffusion $V_s \ll V_0$.

(a)  <u>Vacancy in-diffusion</u>: $V_s \gg V_0$

   Let $C \ll K_2$; this condition may be realized in $Cr,Ca : YAG$ crystals by administering low chromium concentrations $C$, and/or using a high temperature thus increasing the reaction constant $K_2$; for demonstration we chose $C/K_2 = 0.001$. Eq. (16) then reduces to a regular diffusion equation $\partial V/\partial t = D(\partial^2 V / \partial x^2)$. Analytical solution for a host of sample geometrical shapes and boundary conditions are well established, and summarized in books [20,22].

   In Fig. 1 we show the vacancy $V_O^{2-}$, $Cr^{3+}$, $Cr^{4+}$, and reaction rate $R(x,t)$ profiles obtained during reduction of a $Cr,Ca : YAG$ slab. Initially, the sample was practically fully oxidized (mostly $Cr^{4+}$ present - $V_0/K_2 = 0.01$). Reduction is performed under a low ambient oxygen pressure, specifically $V_s/K_2 = 100$. The times involved are such that $\sqrt{Dt} \ll \ell$, where $2\ell$ is the slab thickness. The expressions used in calculating the vacancy diffusion are provided in Appendix A. The $Cr^{4+}$ concentration reduces, starting near the sample surfaces ($x = \pm \ell$), while the $Cr^{3+}$ accordingly increases, as their sum is a constant (Eq. (7)). The reaction rate profile shows a distinct peak, at a distance of approximately $4\sqrt{Dt}$ from the surface, and approximately $\sqrt{Dt}$ wide (full width at half maximum). The zone surrounding the reaction rate peak position is often referred to as the "reaction zone" or the "reaction front" [17,18,19]. Note, that the units of $R$ chosen in Fig. 1, $K_2/t$, are time-dependent, namely, they turn smaller as time progresses. In fact, $R$ itself is thus inversely proportional to time ($R \sim 1/t$). The physical image is that of a reaction front advancing from the surface into the crystal bulk, turning wider and smaller with time.

In contrast to Fig. 1, in Fig. 2 we chose time-independent units for the position and reaction rate axes. Here we show the progress in time of the reaction rate profile when the sample reduction takes place through both end surfaces. With time, the reaction rate peaks move closer to the sample center, simultaneously becoming lower. Eventually the peaks "collide", and diminish asymptotically, as virtually all $Cr^{4+}$ ions reduce into $Cr^{3+}$ ones.

Let $C >> K_2$ (see Appendices B and C); Such conditions may be realized in $Cr, Ca : YAG$ crystals by administering relatively high chromium concentrations $C$, and/or using a low temperature, thus reducing the reaction constant $K_2$. In Figs. 3 and 4 we provide the $[V_O^{2-}]$, $[Cr^{3+}]$, $[Cr^{4+}]$, and reaction rate $R(x,t)$ profiles obtained during reduction of a $Cr, Ca : YAG$ slab under different initial conditions. As in Figs. 1 and 2, the sample is initially almost fully oxidized ($V_0/K_2 = 0.01$), and reduction is performed under a low ambient oxygen pressure yielding $V_s/K_2 = 100$. However, the maximal chromium ion concentration differs grossly in Figs. 3 and 4, as compared to Figs. 1 and 2: $C/K_2 = 200$ in Fig. 3, and $C/K_2 = 10^5$ in Fig. 4. It is clearly seen, that as the maximal chromium concentration grows, the reaction peak position moves closer to the surface:

$(x_{peak} / \sqrt{Dt}) \sim 3.8$ for $C/K_2 = 0.001$; $(x_{peak} / \sqrt{Dt}) \sim 0.55$ for $C/K_2 = 200$;

$(x_{peak} / \sqrt{Dt}) \sim 5 \times 10^{-4}$ for $C/K_2 = 10^5$. Simultaneously, the reaction peak-width turns

narrower: the Full Width at Half Maximum is $FWHM \sim 1.6\sqrt{Dt}$ for $C/K_2 = 0.001$;

$FWHM \sim 0.1\sqrt{Dt}$ for $C/K_2 = 200$; $FWHM \sim 7 \times 10^{-5} \sqrt{Dt}$ for $C/K_2 = 10^5$.

One may note, that in our model, the reaction peak is always obtained at some distance

from the surface. This result is directly related to the assumption of steady-state concentration

at the surface, starting at $t = 0$. In reality, the establishment of a steady state surface

concentration takes some time, during which, the reaction peak position is obviously closer to

the surface. It should be valuable to establish experimentally to what extent the above

assumption is valid; for example, by measuring the surface crossing velocity as described in

[21].

(b) <u>Vacancy out-diffusion</u>: $V_s \ll V_0$

We consider now the reverse case of vacancy out-diffusion. The starting condition is a slab almost fully reduced, containing mostly $Cr^{3+} / V_O^-$ species. Oxidation enforces vacancy out-diffusion, transforming the $Cr^{3+} / V_O^-$ species into $Cr^{4+}$ ones. Transformation starts initially near the surface, and progresses towards the slab center. The $[V_O^{2-}]$, $[Cr^{3+}]$, $[Cr^{4+}]$, and the reaction rate $R(x,t)$ profiles for the oxidation case are shown in Figs. 5 and 6 for relatively short times such that $\sqrt{Dt} \ll \ell$. Fig 5 relates to $C/K_2 \ll 1$, while Fig. 6 relates to

$C/K_2 \gg 1$. Note that in both cases, the reaction rate is negative! Physically this means that

$Cr^{3+} / V_O^-$ species transform into $Cr^{4+}$ ones:- "right to left" in the reaction described by Eq.

(2). In both cases, the reaction is concentrated near the surface; changes in the $[Cr^{3+}]$, and $[Cr^{4+}]$ are limited to the near-surface region. The $[Cr^{3+}]$ is virtually zero at the surface, grows and saturates at the initial concentration value far from it.

In Figs. 7 and 8 we show the same characteristics as in Fig. 5 ($C/K_2 \ll 1$), however for much longer times. Fig. 7 concentrates on the $[Cr^{4+}]$ profiles. At the shorter time $t_1 = 0.02\,(\ell^2/D)$, the $Cr^{4+}$ concentration is still largest near the surface, and turns very small towards the center region. As time progresses, ($t_2 = \ell^2/D$), the $Cr^{4+}$ concentration grows at the center region, while that near the surface remains equal to the steady-state concentration $C_s = 10^{-3} K_2$ imposed by the boundary conditions (see also Eq. (9)). Concentration growth continues at the center region, approaching the asymptotic constant value $10^{-3} K_2$ throughout the sample. Fig. 8 shows the reaction rate profiles $R(x, t)$ for the same process described in Fig. 7. The reaction rate exhibits a peak near the surface for the shorter time; as time progresses, the reaction rate diminishes near the surface, while it is growing in the center region. Eventually, as the $Cr^{4+}$ concentration approaches the asymptotic constant value throughout the sample, the reaction rate diminishes throughout.

The different regions of $C/K_2$ for which the above analyses apply, are demonstrated in Fig. 9. Appendices A, B, and C below relate to the same regions.

It is thus seen, that the dynamic behavior of the system during oxidation is not symmetric with that occuring during reduction. In the case of reduction, a well defined reaction front establishes. Overlooking fine details, in the case of oxidation, the reaction extends through the entire volume of the sample. This qualitative conclusion may be verified by experiment, by studying the cross shape of the sample coloration change during oxidation and during reduction. Similar effects were recently observed in oxidation and reduction of some other oxide crystals [23,24,25].

## V. Comparison with experiment

   In Fig. 10 we show the total concentration of $Cr^{4+}$ ions, octahedral plus tetrahedral, as calculated from the experimental data published by Feldman et al [8], as a function of oxidation time. The Cr:Ca:YAG crystal was in the form of a thin slab, 7.25 mm thick. It was initially fully reduced in vaccum, and then oxidized by exposure to free air at $1000^{o}C$.

Calculations are based on the absorption peak intensities of octahedral and tetrahedral $Cr^{4+}$

ions given in [8, Fig. 5], the oscillator strengths in [8, Table 1], and the relation  between

concentration and oscillator strength in [8, Eq. (A.2)]. The total $Cr^{4+}$ ion concetration grows

with time for short oxidation times ($10 - 280$ min ), and saturates afterwards. In fact, for all

conditions that we have considered for the oxidation process, the $Cr^{4+}$ ion concentration

follows a similar behavior. Using these experimental resuls solely, would not provide any

conclusion on the conditions that the experimental resuts relate to. We however wish to point

out that quite interestingly, the data on the short time region follows the relation

$$\left[Cr^{4+}\right][cm^{-3}] = 1.23 \times 10^{17} (t[min])^{0.66} .$$ Analyses indicate that a similar behavior is

obtained over relatively large ranges in time for two opposite limiting cases: $C/K_2 \ll 1$

[Appendix A], and $C/K_2 \gg 1$, [26]. More experimental work is needed to establish a comprehensive agreement between experiment and theory.

## VI. Conclusions and Summary

   Our present theoretical work relates to experiments conducted by Feldman et al [8] on the dynamics of chromium ion valence transformations in Cr,Ca:YAG crystals among the trivalent $Cr^{3+}$ state and two tetravalent $Cr^{4+}$ ones, of octahedral and tetrahedral coordinations. The basic effects observed were transitions of $Cr^{3+}$ into $Cr^{4+}$ under high-temperature annealing in an oxidizing atmosphere, and the reverse transition under a reducing atmosphere, or under vaccum. We interpret the processes described by oxygen-vacancy diffusion in the YAG bulk. We wrote down the quasi-chemical reaction

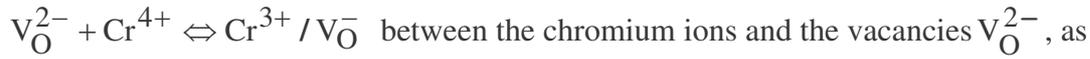 between the chromium ions and the vacancies $V_O^{2-}$, as responsible for the valence transformations. The external oxygen pressure determines the equilibrium oxygen vacancy concentration at the surface, which extends by diffusion through the entire sample after a sufficiently long time. The $Cr^{4+}$ equilibrium concentration turns out to be proportional to the square-root of the oxygen partial pressure at low pressures, and to saturate at high pressures, when the entire $Cr^{3+}$ ion population has oxidized. This conclusion calls for verification by a straight-forward experimental study. Our diffusion equation is based on the assuption of a local equilibrium in the quasi-chemical reaction among the ions and the oxygen vacancies. The present theory provides spatial profiles of the different chromium ion variations throughout the crystal during oxidation and reduction. Reduction is always chracterized by two well-defined zones, located near the sample opposite surfaces, where the $Cr^{4+} \rightarrow Cr^{3+}$ reaction transition takes place. These zones move continually towards the sample center. During oxidation, oxygen vacancies diffuse towards the surface. No reaction front establishes; For sufficiently long times, the $Cr^{3+} \rightarrow Cr^{4+}$ transition reaction progresses throughout the entire sample. A comparison with existing results on the integrated $Cr^{4+}$ concentration as a function of time during oxidation [8] shows a reasonable agreement.

## Appendix A

In the limit $C/K_2 << 1$, Eq. (16) reduces to

$$\frac{\partial V}{\partial t} = D \frac{\partial^2 V}{\partial x^2} \quad , \tag{A1}$$

with the boundary conditions $V(x = \pm \ell) = V_s$ for $t \geq 0$, and initial conditions

$V(t = 0) = V_0$ for $-\ell < x < +\ell$. The solution is provided in [20, p. 48] as

$$\frac{V - V_0}{V_s - V_0} = 1 - \frac{4}{\pi} \sum_{n=0}^{\infty} \frac{(-1)^n}{2n+1} \exp\left\{ -D(2n+1)^2 \pi^2 t / 4\ell^2 \right\} \cos \frac{(2n+1)\pi x}{2\ell} \quad . \tag{A2}$$

A solution useful for times short compared to $\sim \ell^2 / D$ is [20, p 49]

$$\frac{V - V_0}{V_s - V_0} = \sum_{n=0}^{\infty} (-1)^n \ \mathbf{erfc} \ \frac{(2n+1)\ell - x}{2\sqrt{(Dt)}} + \sum_{n=0}^{\infty} (-1)^n \ \mathbf{erfc} \ \frac{(2n+1)\ell + x}{2\sqrt{(Dt)}} \tag{A3}$$

## Appendix B

To address the diffusion problem in the range $1 << C/K_2 << 10^4$, Eqs. (11)-(14) in the

quasi-stationary approximation may be written in the form

$$\frac{\partial V(x,t)}{\partial t} = D \frac{\partial^2 V(x,t)}{\partial x^2} + \frac{\partial [Cr^{4+}]}{\partial t} \quad , \tag{B1}$$

$$[Cr^{4+}] + [Cr^{3+}] = C \quad , \tag{B2}$$

$$\frac{V(x,t) \cdot [Cr^{4+}]}{[Cr^{3+}]} = K_2 \quad , \tag{B3}$$

$$R(x,t) = -\frac{\partial [Cr^{4+}]}{\partial t} \quad . \tag{B4}$$

This form allows to obtain straightforwardly the desired approximate solution in the said range [16]. The boundary conditions are $V(x = \pm \ell) = V_s$ for $t \geq 0$, and the initial conditions are $V(t=0) = V_0$ for $-\ell < x < +\ell$. Defining $x_f(t) \equiv 2\sqrt{D_f t}$, where $D_f$ is the solution of

$$\frac{\sqrt{D/\pi D_f}\ \exp(-D_f/D)}{\mathbf{erf}\ \sqrt{D_f/D}} = \frac{C}{V_s}\ ,$$   (B5)

one obtains the following solution of Eqs (B1)-(B4), for $V_s \gg V_0$:

for $x < x_f(t)$:

$$[Cr^{3+}] = C\ ;\ [Cr^{4+}] = 0\ ;\ V = V_s\left(1 - \frac{\mathbf{erf}(x/2\sqrt{Dt})}{\mathbf{erf}(x_f(t)/2\sqrt{Dt})}\right)\ ;$$   (B6)

for $x > x_f(t)$:

$$[Cr^{3+}] = 0\ ;\ [Cr^{4+}] = C\ ;\ V = 0\ ;$$   (B7)

The $Cr^{4+}$ conversion into $Cr^{3+}$ takes place in the region surrounding $x_f(t)$. Introducing the expanded-scale variable $\xi \equiv (C/K_2)(x - x_f(t))/(4Dt)^{1/2}$, the solution in that region is

$$(V/K_2) + \mathbf{ln}(V/K_2) + (\mathbf{ln}\,2 - 0.5) = -2(D_f/D)^{1/2}\xi\ ;$$   (B8)

$$[Cr^{3+}] = CV/(V + K_2)\ ;$$   (B9)

$$[C^{4+}] = CK_2/(V + K_2)\ ;$$   (B10)

$$R(\xi) = (K_2/t)(D_f/D)C^2 V/(V + K_2)^3.$$   (B11)

The free number $(\mathbf{ln}\,2 - 0.5)$ in (B8) is in fact an integration constant assuring that the reaction rate $R(\xi)$ assumes its maximum at $\xi = 0$, namely at the "Reaction front" $x_f(t)$. One may further note that the condition $C/K_2 \gg 1$ also implies that the typical reaction zone width, $\sim (K_2/C)\sqrt{Dt}$, is small compared to the typical diffusion length $\sim \sqrt{Dt}$.

## Appendix C

In the limit $C/K_2 >> 10^4$, Eq. (16) reduces to

$$\left\{ \frac{C/K_2}{(1 + V/K_2)^2} \right\} \frac{\partial V}{\partial t} = D \frac{\partial^2 V}{\partial x^2} \quad , \tag{C1}$$

whence

$$\frac{\partial}{\partial t} \left[ Cr^{4+} \right] = \frac{\partial}{\partial x} \left( D_{eff} \frac{\partial}{\partial x} \left[ Cr^{4+} \right] \right) \quad , \tag{C2}$$

with

$$D_{eff} \left( [Cr^{4+}] \right) \equiv \frac{D^*}{\left( [Cr^{4+}]/[Cr^{4+}]_s \right)^2} \quad ; \quad D^* \equiv D \frac{CK_2}{[Cr^{4+}]_s^2} \quad ; \tag{C3}$$

$[Cr^{4+}]_s$ is the surface concentration of $Cr^{4+}$, calculated by Eq. (9). An exact analytical

parametric solution is provided in [27] as follows

$$[Cr^{4+}]/[Cr^{4+}]_s = \frac{1}{1 - A\sqrt{\pi} \, \mathbf{exp}(\lambda^2) \, \mathbf{erfc}(\lambda) \left[ 1 - \dfrac{\mathbf{erfc}(\lambda)}{\mathbf{erfc}(A)} \right]} \quad ; \tag{C4 i}$$

$$\frac{x}{\sqrt{D^* t}} = 2 \left\{ \lambda - A \, \mathbf{exp}(A^2 - \lambda^2) - \lambda A \pi^{1/2} \, \mathbf{exp}(A^2) \left[ \mathbf{erfc}(A) - \mathbf{erfc}(\lambda) \right] \right\}, \tag{C4 ii}$$

where $A$ is the solution of the equation

$$A\sqrt{\pi} \, \mathbf{exp}(A^2) \, \mathbf{erfc}(A) = \frac{V_s - V_0}{K_2 + V_0} \quad , \tag{C4 iii}$$

where $\lambda$ is a free real parameter. One may note that in the case $\lambda = A$ corresponds to $x = 0$.

# Figure captions

<u>Fig. 1</u>: $[V_O^{2-}]$, $[Cr^{3+}]$, $[Cr^{4+}]$ and reaction rate $R(x,t)$ profiles obtained during reduction of a $Cr,Ca:YAG$ slab. Sample initially almost fully oxidized ($V_0/K_2 = 0.01$). Reduction performed under a low ambient oxygen pressure, yielding $V_s/K_2 = 100$. Total chromium ion concentration $C/K_2 = 0.001$.

<u>Fig. 2</u>:  Reduced reaction rate profiles for different times after the onset of reduction taking place through both end surfaces:- $t_1 = 0.0067\,(\ell^2/D)$; $t_2 = 0.02\,(\ell^2/D)$;

$t_3 = 0.04\,(\ell^2/D)$; $t_4 = 0.05\,(\ell^2/D)$; $t_5 = 0.1\,(\ell^2/D)$. Initial vacancy concentration

$V_0/K_2 = 0.01$; vacancy surface concentration is $V_s/K_2 = 100$; Total chromium ion

concentration is $C/K_2 = 0.001$.

<u>Fig. 3</u>: $[V_O^{2-}]$, $[Cr^{3+}]$, $[Cr^{4+}]$ and reaction rate $R(x,t)$ profiles obtained during reduction of a $Cr,Ca:YAG$ slab. Sample is initially almost fully oxidized ($V_0/K_2 = 0.01$). Reduction is performed under a low ambient oxygen pressure, yielding $V_s/K_2 = 100$. Total chromium ion concentration is $C/K_2 = 200$.

<u>Fig. 4</u>: $[V_O^{2-}]$, $[Cr^{3+}]$, $[Cr^{4+}]$ and reaction rate $R(x,t)$ profiles obtained during reduction of a $Cr,Ca:YAG$ slab. Sample is initially almost fully oxidized ($V_0/K_2 = 0.01$). Reduction is performed under a low ambient oxygen pressure, yielding $V_s/K_2 = 100$. Total chromium ion concentration is $C/K_2 = 10^5$.

<u>Fig. 5</u>: $[V_O^{2-}]$, $[Cr^{3+}]$, $[Cr^{4+}]$ and reaction rate $R(x,t)$ profiles obtained during oxidation of a $Cr,Ca:YAG$ slab. Sample is initially almost fully reduced ($V_0/K_2 = 100$). Oxidationtion is performed under a high ambient oxygen pressure, yielding $V_s/K_2 = 0.01$. Total chromium ion concentration is $C/K_2 = 10^{-3}$.

<u>Fig. 6</u>: $[V_O^{2-}]$, $[Cr^{3+}]$, $[Cr^{4+}]$ and reaction rate $R(x,t)$ profiles obtained during oxidation of a $Cr,Ca:YAG$ slab. Sample is initially almost fully reduced ($V_0/K_2 = 100$). Oxidationtion is performed under a high ambient oxygen pressure, yielding $V_s/K_2 = 0.01$. Total chromium ion concentration is $C/K_2 = 10^{+5}$.

<u>Fig. 7</u>: $[Cr^{4+}]$ profiles obtained during oxidation of a $Cr,Ca:YAG$ slab. Sample is initially almost fully reduced ($V_0/K_2 = 100$). Oxidationtion is performed under a high ambient oxygen pressure, yielding $V_s/K_2 = 0.01$. Total chromium ion concentration is $C/K_2 = 10^{-3}$. Oxidation times: $t_1 = 0.02\,(\ell^2/D)$; $t_2 = (\ell^2/D)$; $t_3 = 2\,(\ell^2/D)$; $t_4 = 3\,(\ell^2/D)$; $t_5 = 4(\ell^2/D)$.

<u>Fig. 8</u>: Reaction rate profiles $R(x,t)$ obtained during oxidation of a $Cr,Ca:YAG$ slab. Sample is initially almost fully reduced ($V_0/K_2 = 100$). Oxidationtion is performed under a high ambient oxygen pressure, yielding $V_s/K_2 = 0.01$. Total chromium ion concentration is $C/K_2 = 10^{-3}$. Oxidation times: $t_1 = 0.02\,(\ell^2/D)$; $t_2 = (\ell^2/D)$; $t_3 = 2(\ell^2/D)$; $t_4 = 3\,(\ell^2/D)$; $t_5 = 4(\ell^2/D)$

Fig. 9 : Illustration of three regions in the $C/K_2$ - $V/K_2$  plane for which approximate

analytical solutions of Eq. (16) are presented in appendices A, B, and C.

Fig. 10 :  Log-log plot of total $Cr^{4+}$ concentration (octahedral plus tetrahedral) calculated

from the experimental data published in [8], as a function of oxidation time for an initially fully reduced Cr:Ca:YAG crystal, 7.25 mm thick, in an ambient of free air at $1000^{o}C$.

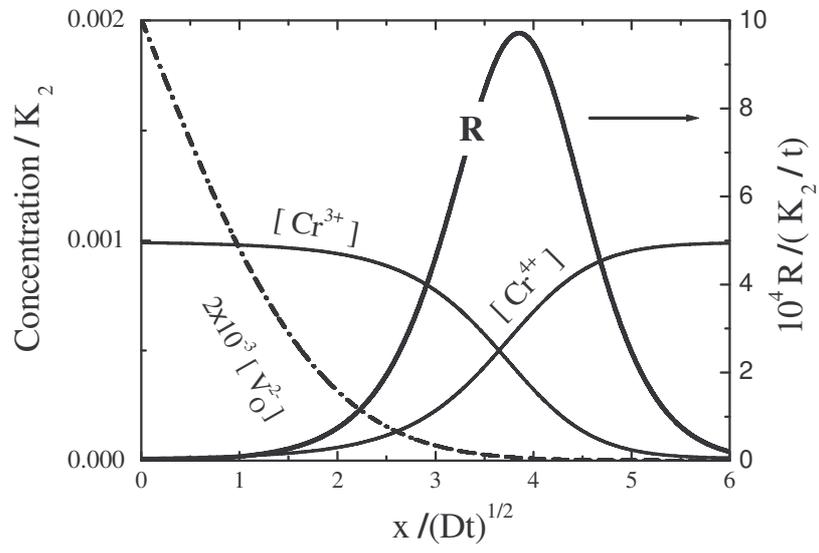

Sinder et al

Fig. 1

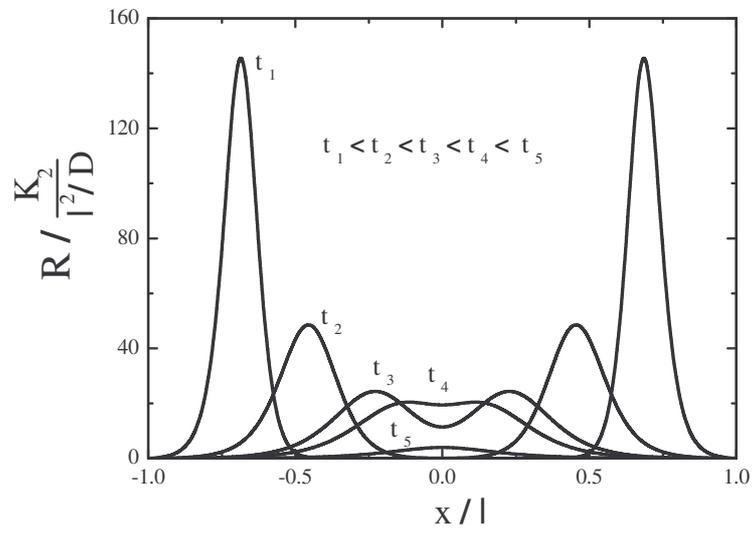

Sinder et al

Fig. 2

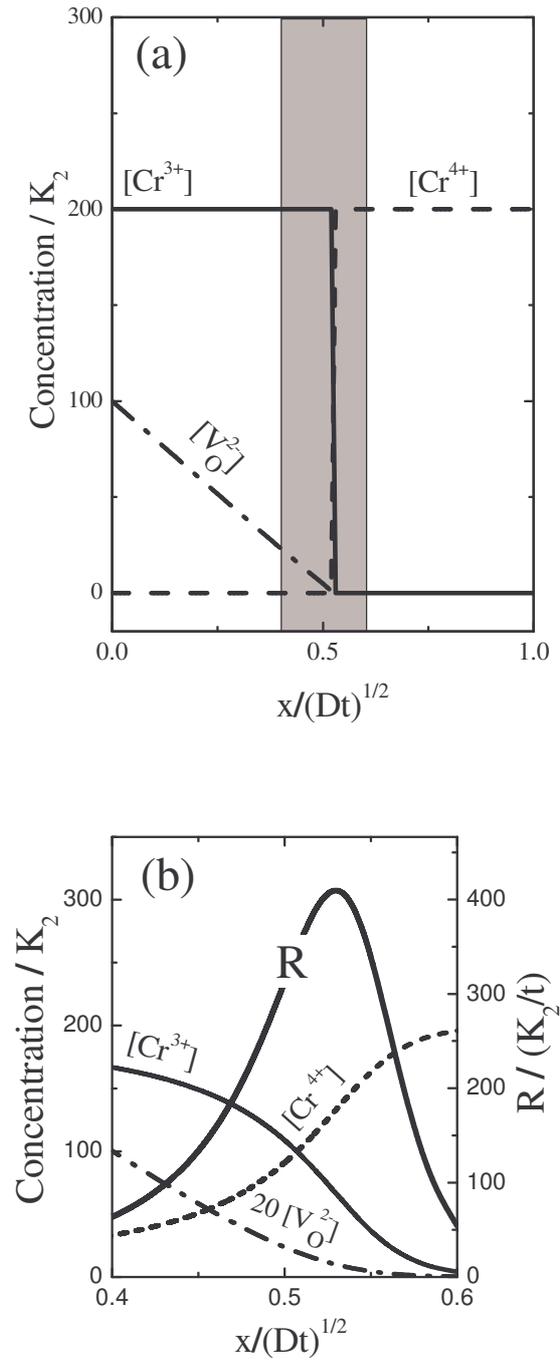

Sinder et al

Fig. 3

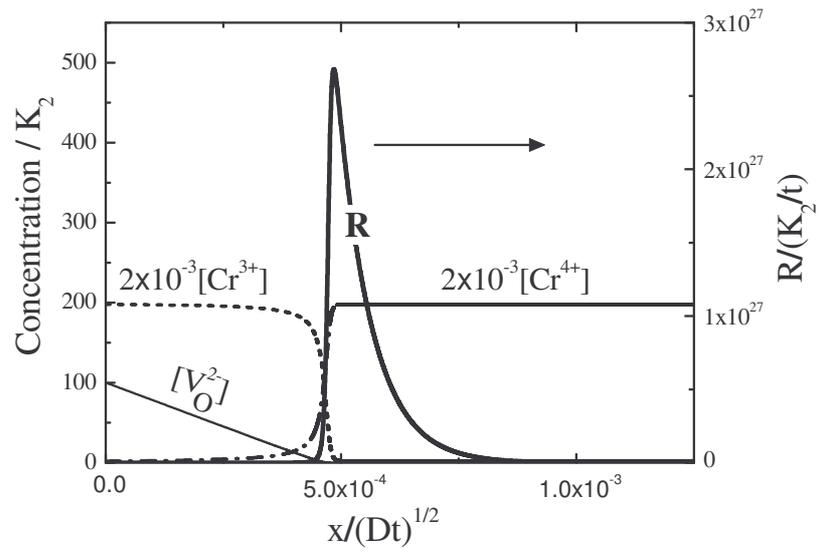

Sinder et al

Fig. 4

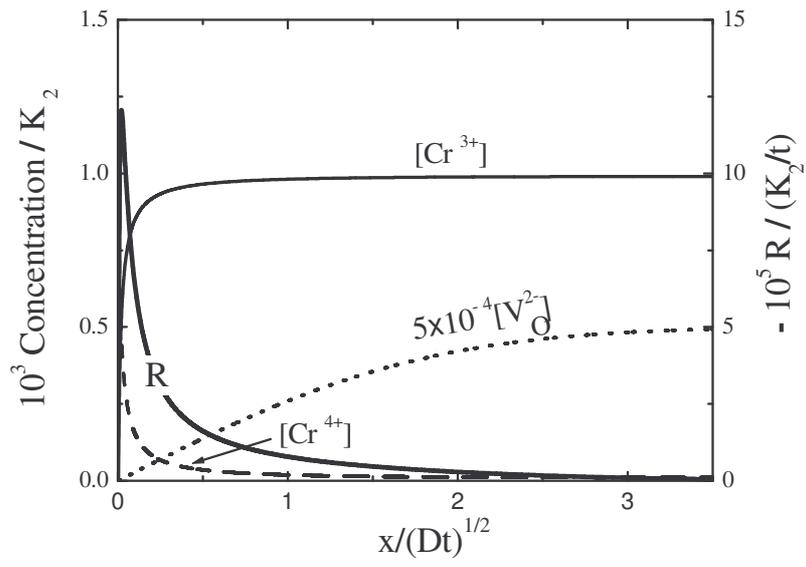

Sinder et al

Fig. 5

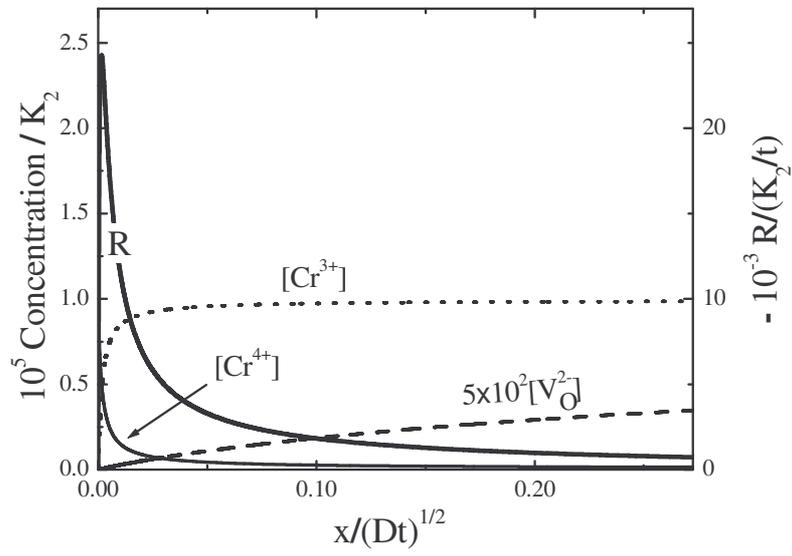

Sinder et al

Fig. 6

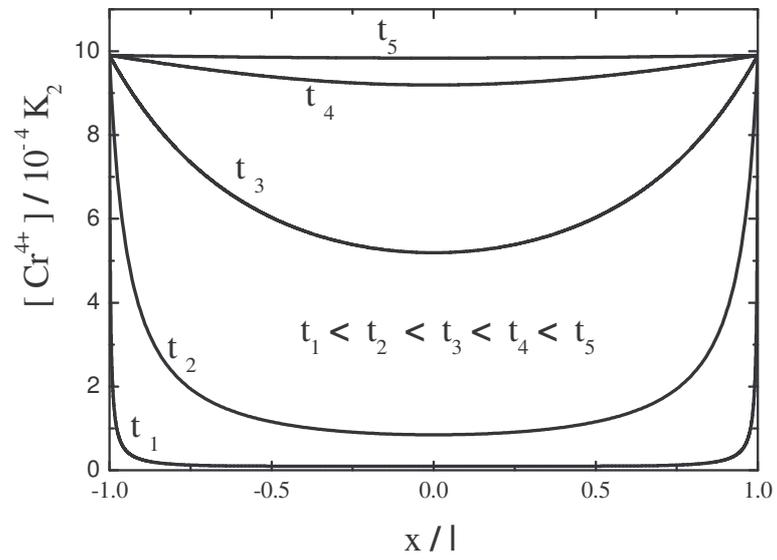

Sinder et al

Fig. 7

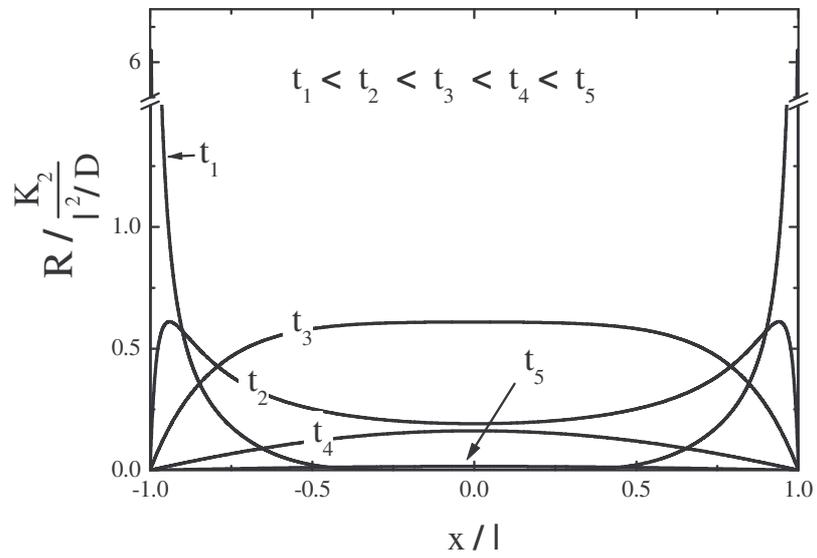

Sinder et al

Fig. 8

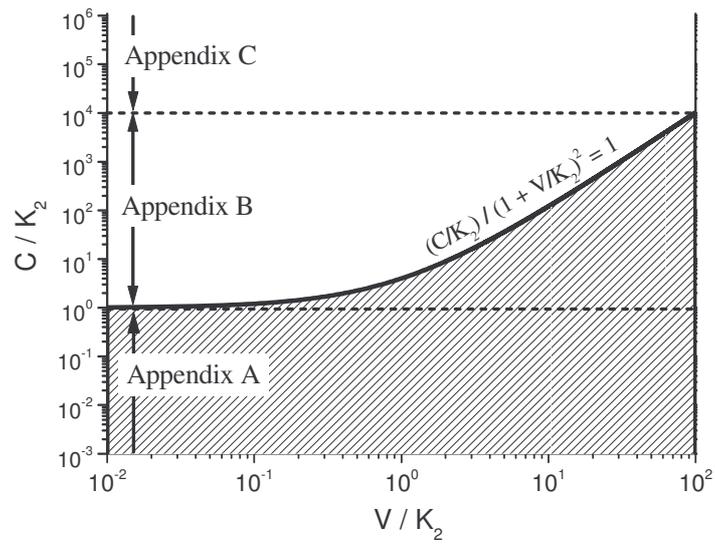

Sinder et al

Fig. 9

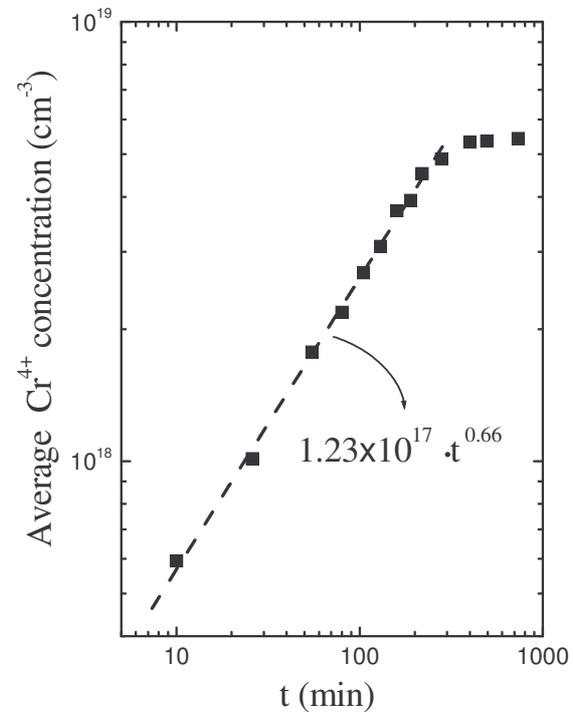